\documentclass{article}
\usepackage{graphicx}
\graphicspath{ {images/} }
\usepackage{amsmath,amssymb,amsfonts}

\renewcommand{\theequation}
{\arabic{section}.\arabic{equation}}

\makeatletter
\def\eqnarray{ \stepcounter{equation} \let\@currentlabel=\theequation
 \global\@eqnswtrue
 \global\@eqcnt\z@
 \tabskip\@centering
 \let\\=\@eqncr
 $$\halign to \displaywidth\bgroup\@eqnsel\hskip\@centering
 $\displaystyle\tabskip\z@{##}$&\global\@eqcnt\@ne
 \hfil$\displaystyle{{}##{}}$\hfil
 &\global\@eqcnt\tw@$\displaystyle\tabskip\z@{##}$\hfil
 \tabskip\@centering&\llap{##}\tabskip\z@\cr}
\makeatother

\makeatletter
\def\@arrayacol{\edef\@preamble{\@preamble \hskip .5\arraycolsep}}
\def\array{\let\@acol\@arrayacol \let\@classz\@arrayclassz
\let\@classiv\@arrayclassiv \let\\\@arraycr\def\@halignto{}\@tabarray}
\makeatother

\makeatletter
\newcounter{subeqncnt}
\def\thesubeqncnt{\alph{subeqncnt}}
\def\subequations{\begingroup%
   \stepcounter{equation}\edef\@tempa{\theequation}%
   \let\c@equation\c@subeqncnt\c@subeqncnt\z@
   \edef\theequation{\@tempa\noexpand\thesubeqncnt}}

\makeatother

\newcommand{\be}{\begin{equation}}
\newcommand{\ee}{\end{equation}}

\newcommand{\beqa}{\begin{eqnarray}}
\newcommand{\eeqa}{\end{eqnarray}}
\newcommand{\nn}{\nonumber}

\begin{document}

\setlength{\baselineskip}{7mm}
\begin{titlepage}
\begin{flushright}

{\tt NRCPS-HE-28-2014} \\

\end{flushright}

\begin{center}
{\Large ~\\{\it  Tensor Gluons \\
and  \\
Proton Structure\footnote{Dedicated to Professor Andrey Slavnov at the occasion of his
75th birthday.}
\vspace{1cm}

}

}

\vspace{1cm}

{\sl George Savvidy

\bigskip
\centerline{${}$ \sl Institute of Nuclear and Particle Physics}
\centerline{${}$ \sl Demokritos National Research Center, Ag. Paraskevi,  Athens, Greece}
\bigskip

}
\end{center}
\vspace{30pt}

\centerline{{\bf Abstract}}

In a recent article we were considering a possibility that inside a proton and, more generally,
inside hadrons there could be additional partons - tensor-gluons, which carry a part
of the proton momentum.  Tensor-gluons have zero electric charge, like gluons, but have a larger spin.
Therefore we call them  tensor-gluons. The nonzero density of tensor-gluons can be generated
by the emission of tensor-gluons by gluons.  Tensor-gluons can further split into the pairs
of tensor-gluons through a different channels.
To describe all these processes one should know the general splitting probabilities for tensor-gluons.
These probabilities should fulfill very general symmetry relations, which we were able to
resolve  by introducing a splitting index.  This approach allows to find out the general
form of the splitting  functions, to derive corresponding DGLAP evolution equations and
to calculate the one-loop Callan-Simanzik beta function for tensor-gluons of a given  spin.
Our results provide a nontrivial consistency check of the  theory and of the Callan-Simanzik beta function
calculations, because the theory has a unique coupling constant and its high energy behavior
should be universal for all particles of the spectrum.   We argue that the contribution of all spins
into the beta function  vanishes leading to a conformal invariant theory at very high energies.

\vspace{12pt}

\noindent

\end{titlepage}



\pagestyle{plain}

\section{\it Introduction}

In the recent article \cite{Savvidy:2014hha} we were considering a possibility that inside a proton and,
more generally,
inside hadrons there could be
additional partons - tensor-gluons, which could carry a part of the proton momentum.  Tensor-gluons
have zero electric charge, like gluons, but have a larger spin. Therefore we call them  tensor-gluons.
The nonzero density of tensor-gluons can be generated by the emission of tensor-gluons
by gluons \cite{Savvidy:2005fi,Savvidy:2005zm,Savvidy:2005ki,Savvidy:2010vb,Georgiou:2010mf,Antoniadis:2011rr}.
The process of gluon splitting into tensor-gluons suggests that part of the proton momentum
which was carried by neutral partons is shared between vector and tensor gluons.

To describe this process one should know the splitting amplitudes of gluon into tensor-gluons.
The tree level scattering amplitudes describing a fusion of gluons into tensor-gluons
were found in \cite{Georgiou:2010mf}.
They are generalizations of the Parke-Taylor scattering amplitude to the case
of two tensor gauge bosons of spin $s$ and $(n-2)$ gluons of spin 1 and allow to extract
the splitting amplitudes \cite{Antoniadis:2011rr}.
With these  splitting amplitudes in hand one can derive the generalization
of the DGLAP evolution equations \cite{ Altarelli:1977zs,Dokshitzer:1977sg,
Gribov:1972ri,Gribov:1972rt, Lipatov:1974qm}
for the parton distribution functions, which take into account the processes of
emission of tensor-gluons by gluons \cite{Savvidy:2014hha}.
In this approach  the momentum sum rule allows to find the contribution
of tensor-gluons  of spin s  into the one-loop
Callan-Simanzik beta function of spin 1 gluons, which takes the following form \cite{Savvidy:2014hha}:
\be\label{fullbetaint}
\alpha(t)= {\alpha \over 1+ b  \alpha ~t }~,~~~~~b_{11}=  {\sum_s(12s^2 -1) C_2(G) - 4 n_f T(R) \over 12 \pi},~~~~s=1,2,...,
\ee
where $s$ is the spin of the tensor-gluons running in the loop\footnote{This notation for the coefficient $b_{11}$
aims to show that external particles are spin 1 gluons.}.
The spin-dependent term $12 s^2 -1$ in the Callan-Simanzik beta function coefficient (\ref{fullbetaint})
makes the asymptotic freedom \cite{Gross:1973ju,Gross:1974cs,Politzer:1973fx}
even stronger and influence
the high energy behavior ($t= \ln(Q^2/Q^2_0)$)  of the structure functions.
It also influences the unification scale at which the coupling constants of the
Standard Model merge  \cite{Georgi:1974sy,Georgi:1974yf},
shifting its value to lower energies \cite{Savvidy:2014hha}.

The consistency of the theory requires that the contribution
of the tensor-gluons  into the one-loop
Callan-Simanzik beta function $b_{11}$ for spin 1 gluons should be equal to
the contribution of high spin tensor-gluons to   the
one-loop Callan-Simanzik beta function  $b_{rr}$  for tensor-gluons of spin $r=2,3,...$
This nontrivial  consistency requirement follows from the fact that the theory has unique coupling
constant and its high energy behavior should be universal for all particles of the spectrum.

In order to approach this problem one should know the full set of splitting
probabilities $P^C_{BA}$ which describe the decay of tensor-gluon of any
spin $A=r$  into two tensor-gluons of arbitrary spins $B$ and $C$.
These probabilities were not available \cite{Savvidy:2014hha}.
The splitting probabilities $P^C_{BA}$ should fulfill very
general symmetry relations \cite{ Altarelli:1977zs,Dokshitzer:1977sg,
Gribov:1972ri,Gribov:1972rt, Lipatov:1974qm}, which, as we
shall demonstrate in this article, allow to find the required functions.
The important input information which allows to resolve  the  symmetry
equations  are the splitting functions$P^1_{11}$ and  $P^s_{s1}$ of spin 1 gluon into the tensor-gluons of spins s  which were found earlier in  \cite{Savvidy:2014hha,Antoniadis:2011rr}.

The present paper is organized as follows. In section two the basic formulae for
splitting functions and their symmetry relations are recalled, definitions  and notations
are specified. In section three we introduce the indexes of the splitting functions and extract
them from the known splitting functions $P^1_{11}$ and $P^s_{s1}$. This approach allows to
identify the unknown parameters and to find out the  general structure of the splitting probabilities
$P^s_{s-r \pm 1 ~r}$ of a tensor-gluon of spin r into two tensor-gluons of spins $s-r \pm 1$
and $s$  (\ref{setofgluon2polarizeds-r+1}), (\ref{setofgluon2polarizeds-r-1}).
In section four
the generalized DGLAP evolution equations that include all tensor-gluon splitting
probabilities are derived and the consistency of the one-loop
Callan-Simanzik beta function calculation is demonstrated.   We argue that the contribution of all spins into the beta function  vanishes leading to a conformal invariant theory at very high energies.
In section five we summarise the results.
The Appendix provides the details of the regularization scheme.

\section{\it Interaction Vertices and Splitting Functions}

In the generalized Yang-Mills theory \cite{Savvidy:2005fi,Savvidy:2005zm,Savvidy:2005ki,Savvidy:2010vb}
all interaction vertices
between high-spin particles have {\it dimensionless coupling constants},
which means that the helicities $h_i, i=1,2,3$ of the interacting particles in the vertex are
constrained  by the relation\footnote{The dimensionality
of the three-point  vertex  $M_3(1^{h_1} ,2^{h_2},3^{h_3} )$  is $[mass]^{D=\pm(h_1+h_2+h_3)}$.}
\be
h_1+h_2+h_3= \pm 1~~.
\ee
Therefore {\it on-mass-shell} interaction vertex between
massless tensor-gluons, the $TTT$-vertex,   has the following form \cite{Georgiou:2010mf}:
\beqa\label{dimensionone1}
M_3(1^{h_1} ,2^{h_2},3^{h_3} )~~&=&~~ g f^{abc} <1,2>^{-2h_1 -2h_2 -1} <2,3>^{2h_1 +1} <3,1>^{2h_2 +1},~~~~
h_3= -1 - h_1 -h_2, \nn\\
M_3(1^{h_1} ,2^{h_2},3^{h_3} )~~&=&~~ g f^{abc} [1,2]^{2h_1 +2h_2 -1} [2,3]^{-2h_1 +1} [3,1]^{-2h_2 +1},~~~~~~~~~~~~~~~~~~h_3= 1 - h_1 -h_2,
\eeqa
where g is the YM coupling constant and $f^{abc}$ are the structure constants
of the internal gauge group G\footnote{
In subsequent equations we shall not write the factor $g f^{abc}$ explicitly. It is also understood that in a spinor representation of the on-mass-shell three-particle interaction vertices (\ref{dimensionone1})
the particles momenta are complexly deformed \cite{Dixon:1996wi,Parke:1986gb,Witten:2003nn,
Britto:2005fq,Benincasa:2007xk,Cachazo:2004kj,ArkaniHamed:2008yf,Berends:1988zn,Mangano:1987kp}.}.
In particular, considering the interaction between a gluon of spin $A=1$ (helicities $h_2 = \pm 1$) and
a tensor-gluon of spin $C=s$ (helicities $h_1 = \pm s$), the gluon-tensor-tensor GTT-vertex,
one can find from  (\ref{dimensionone1})  three solutions for the helicities of the third particle
\be\label{vertecies}
h_3 = \pm \vert s-2 \vert,~ \pm s,~
 \pm \vert s+2 \vert~,
\ee
thus its spin can take three different values $B=s-2, s, s+2$.
A simple analysis shows that all possible vertices can be counted from (\ref{vertecies}),
if one considers two cases:
$B =   s-2 $, $s \geq 3$
and $B =  s$, $s \geq 1$. The corresponding  interaction vertices GTT have
therefore the following form:
\beqa\label{1ssvertex}
M_3(1^{-s} ,2^{+1},3^{s-2} )&\propto &   {<1,3>^{4} \over <1,2> <2,3> <3,1>}
\left({<1,2>  \over  <2,3> }\right)^{2s-2},~~~~s \geq 3\nn\\
M_3(1^{+s} ,2^{-1} ,3^{-s} )&\propto&   {<2,3>^{4} \over <1,2> <2,3> <3,1>}
\left({<2,3>  \over  <1,2> }\right)^{2s-2},~~~~s \geq 1.
\eeqa
The next type of interaction vertex involves tensor-gluons of spin $A=2$ (helicities $h_2=\pm 2$)
and of spin $C=s$ (helicities $h_1 = \pm s$).   In this case one can find from  (\ref{dimensionone1})
four solutions:
\be\label{verteciesspin2}
h_3 = \pm \vert s-3 \vert,~ \pm \vert s-1 \vert,~
 \pm \vert s+1 \vert~, \pm \vert s+3 \vert~.
\ee
All possible vertices can be obtained, if one considers  only two
cases: $B =  s-3  $,  $s\geq 5 $ and  $B =  s-1 $ ,  $s \geq 3$.
The corresponding tensor-tensor-tensor  interaction vertices $TTT$ have
the following form:
\beqa\label{2ssvertex}
M_3(1^{-s} ,2^{+2} ,3^{+(s-3)} )&\propto&   {<3,1>^{4} \over <1,2> <2,3> <3,1>}
\left({<1,2>  \over  <2,3> }\right)^{2s-2} \left({<3,1>  \over  <1,2> }\right)^{2},\nn\\
     M_3(1^{+s} ,2^{-2},3^{-(s-1)} )&\propto&   {<2,3>^{4} \over <1,2> <2,3> <3,1>}
\left({ <2,3> \over  <1,2> }\right)^{2s-2}\left({<1,2> \over  <3,1>  }\right)^{2}.
\eeqa
In the general case the interaction vertex, which involves the tensor-gluons
of spin $A=r$ (helicities $h_2=\pm r$) and of spin $C=s$ (helicities $h_1 = \pm s$),
can be of four types (\ref{dimensionone1}):

\be\label{verteciesspinr}
h_3 = \pm \vert s-r-1 \vert,~ \pm \vert s-r+1 \vert,~
 \pm \vert s+r-1 \vert~, \pm \vert s+r+1 \vert~.
\ee
In order to count all possible types of vertices it is sufficient
to consider two cases of spin: $B =  s-r-1   $, $s \geq 2r+1$ and
$B =  s-r+1 $, $s \geq 2r-1$,  while $r=1,2,3...$.
\beqa\label{rssvertex}
M_3(1^{-s} ,2^{+r} ,3^{+(s-r-1)} )&\propto&   {<3,1>^{4} \over <1,2> <2,3> <3,1>}
\left({<1,2>  \over  <2,3> }\right)^{2s-2} \left({<3,1>  \over  <1,2> }\right)^{2r-2},\nn\\
     M_3(1^{+s} ,2^{-r},3^{-(s-r+1)} )&\propto&  {<2,3>^{4} \over <1,2> <2,3> <3,1>}
\left({<2,3> \over   <1,2> }\right)^{2s-2}\left({<1,2> \over  <3,1>  }\right)^{2r-2}.
\eeqa
At $s=1$ and $r=1$ they all reduce to the YM vertex \cite{Savvidy:2014hha}.

Using these vertices  one can compute the scattering amplitudes involving  tensor-gluons
\cite{Georgiou:2010mf}.  They  reduce to the
famous Parke-Taylor formula \cite{Parke:1986gb} when $s=1$ and
can be used to extract splitting amplitudes
of a gluon into two tensor-gluons \cite{Antoniadis:2011rr}.
Considering the amplitudes in the limit when two neighboring particles
become collinear, $k_B  \parallel k_C$, that is,
$k_B = z k_A,~k_C = (1-z) k_A$,  $k^2_A \rightarrow 0$ and $z$ describes the
longitudinal momentum sharing with the corresponding behavior of spinors
$
\lambda_B = \sqrt{z} \lambda_A,~~~\lambda_C = \sqrt{1-z} \lambda_A,
$
one can deduce that the amplitude  takes the  factorization form
\cite{Dixon:1996wi,Parke:1986gb,Berends:1988zn,Mangano:1987kp,Antoniadis:2011rr}:
\be\label{factorization}
M^{tree}_n(...,B^{h_B},C^{h_C},...)~~  {B \parallel C \over \rightarrow}  ~~
\sum_{h_A }
Split^{ tree }_{-h_A}(B^{h_B},C^{h_C})~ \times ~M^{tree}_{n-1}(...,A^{h_A} ,...),
\ee
where $Split^{tree}_{-h_A}(B^{h_B},C^{h_C})$ denotes the splitting amplitude
$A \rightarrow B+C$ and the
intermediate state $A$ has momentum $k_A=k_B +k_C$ and helicity $h_A$.
Since the collinear limits of the scattering amplitudes
are responsible for parton evolution  \cite{Altarelli:1977zs}
one can extract from (\ref{factorization}) the
Altarelli-Parisi splitting probabilities for
tensor-gluons. Indeed, the residue of the collinear
pole in the square (of the factorized
amplitude (\ref{factorization})) gives Altarelli-Parisi splitting probability $P(z)$
\cite{Dixon:1996wi,Parke:1986gb,Berends:1988zn,Mangano:1987kp,Antoniadis:2011rr}:
\be\label{AltarelliParisi}
P^{C}_{BA}(z)= C_2(G) \sum_{h_A , h_B, h_C} \vert Split_{-h_A}(B^{h_B},C^{h_C}) \vert^2 ~ s_{BC},
\ee
where $s_{BC}=2 k_B \cdot k_C= <B,C>[B,C]$.
The invariant operator $C_2$ for the representation R is defined by the equation
$ t^a t^a  = C_2(R)~ 1 $ and $tr(t^a t^b) = T(R) \delta^{ab}$.
The same splitting probabilities can be extracted directly by considering   {\it of-mass-shall}
decay of the  particle A.  It describes the probability of  finding a particle B inside a particle A
with fraction z of the longitudinal momentum of A  and radiation of the third particle C
with fraction $(1-z)$ of the longitudinal momentum of A \cite{Altarelli:1977zs}
\be\label{transverce}
P^{C}_{BA}(z)= {1\over 2} z (1-z) \bar{ \sum}_{helicities}
{\vert M_{A \rightarrow B+C}\vert^2 \over p^{2}_{\perp}},
\ee
where a sum is over the helicities  of B and C and an average over the helicity of A.

\begin{figure}
\includegraphics[width=6cm]{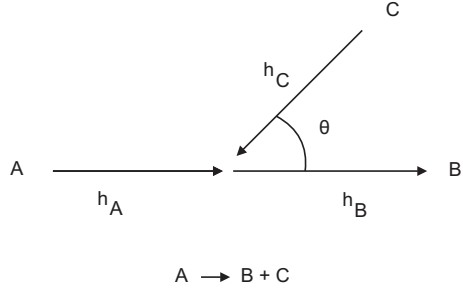}
\centering
\caption{The decay of the tensor-gluon of spin A into
tensor-gluons of spins B and C. The arrows
show the directions of the helicities.
The corresponding splitting function is defined as $P^{C^-}_{B^+A^+}$.  }
\label{fig1}
\end{figure}

It was found that the full set of splitting amplitudes
describing the decay of gluons and tensor-gluons  is \cite{Antoniadis:2011rr}:
\beqa\label{spliting1}
Split_+(B^{+s},C^{-s}) =    \frac{(1-z)^{s+1/2}} {z^{s-1/2}} \frac{1}{ <B, C>},~~~
Split_+(B^{-s},C^{+s}) =    \frac{z^{s+1/2}}{(1-z)^{ s-1/2}} \frac{1}{ <B, C>},\nn\\
Split_{+s}(B^{+},C^{-s}) =  \frac{(1-z)^{s+1}}{\sqrt{z(1-z)}}
\frac{1}{ <B, C>},~~~~~~
Split_{+s}(B^{-s},C^{+}) =  \frac{z^{s+1}}{\sqrt{z(1-z)}}
\frac{1}{ <B, C>},\nn\\
 Split_{-s}(B^{+s},C^{+}) =  \frac{z^{-s+1}}{\sqrt{z(1-z)}}
\frac{1}{ <B, C>}, ~~~~~~
Split_{-s}(B^{+},C^{+s}) =    \frac{(1-z)^{-s+1}}{\sqrt{z(1-z)}}
\frac{1}{ <B, C>}.
\eeqa
This set of splitting amplitudes (\ref{spliting1}) $G\rightarrow TT$, $T \rightarrow GT$ and
$T \rightarrow TG$ reduces to the full set of gluon splitting amplitudes
when $s=1$.
Substituting the splitting amplitudes (\ref{spliting1})
into (\ref{AltarelliParisi}) we are getting \cite{Savvidy:2014hha}:
\beqa\label{setoftensorgluon}
P^T_{TG}(z) &=&  C_2(G)\left[ {z^{2s+1} \over (1-z)^{2s-1}}
+ {(1-z)^{2s+1} \over z^{2s-1}} \right],\nn\\
P^T_{GT}(z) &=&  C_2(G)\left[ {1\over z(1-z)^{2s-1}}
+{(1-z)^{2s+1} \over z} \right],\\
P^G_{TT}(z) &=&  C_2(G)\left[{1 \over (1-z)z^{2s-1}}
+   {z^{2s+1} \over (1-z)}
\right]. \nn
\eeqa
In the leading order the kernel $P^T_{TG}(z)$ has a meaning of variation of the probability
density of finding a tensor-gluon inside the gluon, $P^T_{GT}(z)$ - of finding gluon inside
the tensor-gluon and $P^G_{TT}(z)$ - of finding tensor-gluon inside the tensor-gluon.

The important properties
of the splitting functions are the symmetries \cite{ Altarelli:1977zs,Dokshitzer:1977sg,
Gribov:1972ri,Gribov:1972rt, Lipatov:1974qm} over exchange of the
particles $B \leftrightarrow C$ (see Fig. \ref{fig1} ) with
complementary momenta fraction
\be\label{symmetryrelations1}
P^{C}_{BA}(z)= P^{B}_{CA}(1-z)
\ee
and a crossing relation
\be\label{symmetryrelations2}
P^{C}_{AB}(z)= (-1)^{2h_A +2h_B +1} z P^{C}_{BA}({1\over z}),
\ee
which emerge because two splitting processes are connected
by time reversal $A  \leftrightarrow B$. {\it We shall postulate their validity for interacting
particles of any spins.}
It is easy to see that these relations fulfill in the case of higher spins (\ref{setoftensorgluon}):
\beqa
&P^{T}_{TG}(z)=P^{T}_{TG}(1-z),~~~~~~~~~~~~~~~~~P^{T}_{GT}(z)=P^{G}_{TT}(1-z),~~~~~~~0< z < 1, \nn\\
&P^{T}_{TG}(z)=- z P^{T}_{GT}({1\over z}),~~~~~~~~~~~~~~~P^{G}_{TT}(z)=- z P^{G}_{TT}({1\over z}).~~~~~~~~~~~~~~~~~~
\eeqa
For completeness we shall present also quark and gluon
splitting probabilities \cite{Altarelli:1977zs}:
\beqa\label{setofquarkgluon}
P^G_{qq}(z) &=& C_2(R){1+z^2 \over 1-z },\nn\\
P^q_{Gq}(z) &=& C_2(R)[{1\over z} +{(1-z)^2 \over z }],\\
P^q_{qG}(z) &=& T(R)[z^2 +(1-z)^2], \nn\\
P^G_{GG}(z) &=&  C_2(G)\left[{1 \over z(1-z)}+ {z^4 \over z(1-z)}+{(1-z)^4 \over z(1-z)}\right],\nn
\eeqa
where $C_2(G)= N, C_2(R)={N^2-1  \over  2 N},  T(R) = {1  \over  2}$ for the SU(N) groups.

Let us consider the splitting probabilities (\ref{setoftensorgluon}) in
the limit of the half-integer spin $s \rightarrow 1/2$. One can see  that they
reduce to the quark-gluon splitting probabilities  (\ref{setofquarkgluon})
\be
P_{TT}(z)  \rightarrow P_{qq}(z),~~~P_{GT}(z)  \rightarrow P_{Gq}(z),~~~
P_{TG}(z)  \rightarrow P_{qG}(z)
\ee
and that in the limit $s \rightarrow 1$ to the gluon-gluon splitting probability
\be
{1\over 2}(P_{TG}(z) +P_{GT}(z)  + P_{TT}(z) ) \rightarrow  P_{GG}(z).
\ee
The set of splitting probabilities (\ref{setoftensorgluon}) and (\ref{setofquarkgluon}) provides
an important data for the extension of the results to more general cases of transitions between
high spin tensor, which we shall consider in the next section.

\section{\it General Structure of Splitting Probabilities}

As one can observe from the previous examples the splitting probabilities
are polynomial functions of a simple form:
\be
P^C_{BA} \propto z^n (1-z)^m .
\ee
Using the crossing symmetry relation (\ref{symmetryrelations2}) for the interchange
$A \leftrightarrow  B$ and the fact that we are considering integer spins
\be
P^C_{AB}(z) = (-1)^{2 h_A +2 h_B +1} z P^C_{BA}({1 \over z})=
-z P^C_{BA}({1 \over z}),
\ee
one can find
\be
P^C_{AB}(z) \propto - z {1\over z^n} (1-{1\over x})^m= - {(z-1)^m \over z^{n+m-1}},
\ee
and because the probabilities should be positive it follows that the integer
$m$ should be odd, $m \rightarrow 2m +1$:
\be
P^C_{BA} \propto z^n (1-z)^{2m+1},~~~~~~P^C_{AB} \propto  {(1-z)^{2m+1} \over z^{2m+n}}
\ee
The interchange (\ref{symmetryrelations1}) of particles $A \leftrightarrow  C$ and then of the
$C \leftrightarrow  B$ gives
\be
P^{A}_{CB} \propto  {z^{2m+1} \over (1-z)^{2m+n}}~~\Rightarrow~~~
P^{A}_{BC} \propto -z {z^{2m+n} \over z^{2m+1}(z-1)^{2m+n}}=
-{z^{n} \over (z-1)^{2m+n}}.
\ee
It follows then that the integer $n$ also should be odd, $n \rightarrow 2n+1$, thus the splitting probabilities
should have the following general form:
\be
P^C_{BA} \propto z^{2n+1} (1-z)^{2m+1}  .
\ee
The full set of splitting probabilities can be found by using symmetries
(\ref{symmetryrelations1}), (\ref{symmetryrelations2}) and has the following form:
\beqa\label{generalprobabilities}
&P^C_{BA} \propto z^{2n+1} (1-z)^{2m+1},~~~~
P^{B}_{CA} \propto (1-z)^{2n+1} z^{2m+1} ,\nn\\
&P^{C}_{AB} \propto {(1-z)^{2m+1} \over z^{2n+2m +1}},~~~~~~~~
P^{B}_{AC} \propto {(1-z)^{2n+1} \over z^{2n+2m +1}}, \nn\\
&P^A_{CB} \propto {z^{2m+1} \over (1-z)^{2n+2m+1}},~~~~~~~
P^{A}_{BC} \propto {z^{2n+1} \over (1-z)^{2n+2m+1}}.~~~
\eeqa
In order to identify the above parameters $n$ and $m$ with the helicity content of the interacting particles
let us consider the analytical continuation of these functions to the full
complex plane $C^2$ and take the limit $\vert z \vert \rightarrow \infty$.
We shall define the ratio
\be\label{index}
\lim_{\vert z \vert \rightarrow \infty}~~ {\ln P \over \ln \vert z \vert}= \nu
\ee
as the
{\it index}~ of a given splitting function P, thus from (\ref{generalprobabilities}) we have
\beqa\label{index}
\nu =\lim_{\vert z \vert \rightarrow \infty}~~~~\{
\begin{array}{llll}
&{\ln P^C_{BA}\over \ln \vert z \vert} &\propto +2n+2m +2 ,~~~~~~~~~~~
{\ln P^{B}_{CA}\over \ln \vert z \vert}   &\propto  +2n+ 2m+2  ,\\
&{\ln P^A_{CB}\over \ln \vert z \vert}  &\propto  - 2n ,~~~~~~~~~~~~~~~~~~~~~~~~
{\ln P^{B}_{AC}\over \ln \vert z \vert}  &\propto -2m , \\
&{\ln P^{C}_{AB}\over \ln \vert z \vert} &\propto -2n ,~~~~~~~~~~~~~~~~~~~~~~~~
{\ln P^{A}_{BC}\over \ln \vert z \vert}  &\propto -2m .
\end{array}
\eeqa

\begin{figure}[!]
\includegraphics[width=6cm]{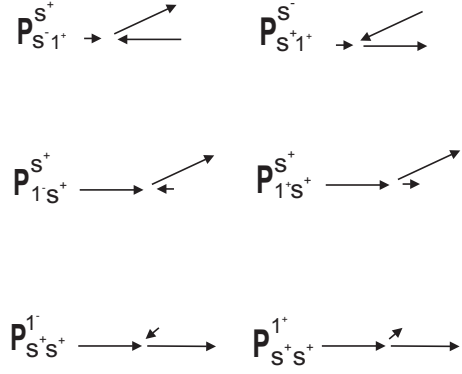}
\centering
\caption{The helicity diagrams  corresponding to the splitting functions
$P^{C \nwarrow }_{B  \leftarrow  A }$
represent  the decay of a parton of  spin $A $ into the partons of spin
$B $  and $C $ in (\ref{setoftensorgluonpolarized}).
The short arrows display spin 1 partons, the  long arrows display  spin s partons.
}
\label{fig2}
\end{figure}

Let us now compare these indexes with the indexes of the known
GTT polarized splitting functions (\ref{spliting1}), (\ref{setoftensorgluon}):
\beqa\label{setoftensorgluonpolarized}
P^{s^+}_{s^- 1^+}(z) &=&  C_2(G) {z^{2s+1} \over (1-z)^{2s-1}},~~~~~
P^{s^-}_{s^+ 1^+}(z) = C_2(G) {(1-z)^{2s+1} \over z^{2s-1}},\nn\\
P^{s^+}_{1^- s^+}(z) &=&  C_2(G) {1 \over z(1-z)^{2s-1}},~~~~~
P^{s^+}_{ 1^+ s^+}(z) = C_2(G) {(1-z)^{2s+1} \over z}, \\
P^{1^-}_{s^+ s^+}(z) &=&  C_2(G) {1  \over (1-z) z^{2s-1}},~~~~~
P^{1^+}_{s^+ s^+}(z) =  C_2(G) { z^{2s+1}  \over (1-z)}, \nn
\eeqa
where $s \geq 1$. We shall get the following identification of the parameters $n$ and $m$ in the
general expression (\ref{generalprobabilities}):
$$
2n+2m+2 =2,~~~~-2n=-2s,~~~~-2m=2s,
$$
and the parameters corresponding to all possible
polarizations of the interacting particles therefore are
\be
A=1,~~ B=s,~~ C=s ~~~ \Rightarrow ~~~ n= +s,~~~m =- s.
\ee
This result has very simple meaning if one considers the diagrams on Fig.\ref{fig2} showing the
distribution of helicities  between interacting particles corresponding to the splitting
functions  (\ref{setoftensorgluonpolarized}). Indeed, because in the generalized YM theory
interaction vertices have
one derivative \cite{Savvidy:2014hha},  the total sum of helicities  during the interactions
can be changed only by one
unit of angular momentum   (\ref{dimensionone1}). Thus when the decaying parton has spin 1 and
outgoing patrons have large spin s
the only way to fulfill the angular momentum conservation (\ref{dimensionone1}) is to direct
helicities of the tensor-gluons in opposite direction, as it is illustrated by the upper diagrams  in Fig.\ref{fig2}.
When initial parton  has a large spin s and decays into patrons of spin 1 and s, the direction
of the large helicities cannot be changed, as it is illustrated by the last four diagrams in Fig.\ref{fig2}.

Our aim now is to derive the splitting probabilities for the
tensor-gluons of spin $A=2$. The corresponding content of the particle spins is
given in (\ref{verteciesspin2}), therefore
\be
A=2,~~B=s-1,~~C=s ,~~~~\Rightarrow~~~~ n= (s-1),~~~m =- (s-1)
\ee
and
\beqa\label{setofgluon2polarizeds-1}
P^{s^+}_{(s-1)^- ~2^+}(z) &=&  C_2(G) {z^{2s-1} \over (1-z)^{2s-3}},~~~~~
P^{(s-1)^-}_{s^+ 2^+}(z) = C_2(G) {(1-z)^{2s-1} \over z^{2s-3}},\nn\\
P^{s^+}_{ 2^- ~(s-1)^+}(z) &=&  C_2(G) {1 \over z(1-z)^{2s-3}},~~~~~
P^{(s-1)^+}_{2^+ s^+}(z) = C_2(G) {(1-z)^{2s-1} \over z}, \\
P^{2^-}_{s^+~ (s-1)^+}(z) &=&  C_2(G) {1  \over (1-z) z^{2s-3}},~~~~~
P^{2^+}_{(s-1)^+~ s^+}(z) =  C_2(G) { z^{2s-1}  \over (1-z)}, \nn
\eeqa
where $s \geq 3$. In the second basic case of spin $B=s-3$ given in (\ref{verteciesspin2}),
\be
A=2,~~B=s-3,~~C=s ~~~~\Rightarrow~~~~ n= (s-3),~~~m =- (s-3),
\ee
the splitting probabilities are:
\beqa\label{setofgluon2polarizeds-3}
P^{s^+}_{(s-3)^- ~2^+}(z) &=&  C_2(G) {z^{2s-5} \over (1-z)^{2s-7}},~~~~~
P^{(s-3)^-}_{s^+ 2^+}(z) = C_2(G) {(1-z)^{2s-5} \over z^{2s-7}},\nn\\
P^{s^+}_{ 2^- ~(s-3)^+}(z) &=&  C_2(G) {1 \over z(1-z)^{2s-7}},~~~~~
P^{(s-3)^+}_{2^+ s^+}(z) = C_2(G) {(1-z)^{2s-5} \over z}, \\
P^{2^-}_{s^+~ (s-3)^+}(z) &=&  C_2(G) {1  \over (1-z) z^{2s-7}},~~~~~
P^{2^+}_{(s-3)^+~ s^+}(z) =  C_2(G) { z^{2s-5}  \over (1-z)}, \nn
\eeqa
where $s \geq 5$.
For general spins  $A=r$ and $C=s$ the third spin takes the values (\ref{verteciesspinr})
\beqa\label{verteciesspinrr}
&&B =  s-r+1  ,~~~~~~ s \geq 2r-1,\nn\\
&&B =  s-r-1  ,~~~~~~ s \geq 2r+1,~~~~~~r=1,2,3,.....
\eeqa
In these cases we shall get that
\beqa
A=r,~~~B=s-r+1,~~~~C=s~~~~~ \Rightarrow~~~~ n= (s-r+1),~~~m =- (s-r+1)\nn,
\eeqa
and the splitting probabilities are
\beqa\label{setofgluon2polarizeds-r+1}
P^{s^+}_{(s-r+1)^- ~r^+}(z) &=&  C_2(G) {z^{2s-2r+3} \over (1-z)^{2s-2r+1}},~~~~~
P^{(s-r+1)^-}_{s^+ r^+}(z) = C_2(G) {(1-z)^{2s-2r+3} \over z^{2s-2r+1}},\nn\\
P^{s^+}_{ r^- ~(s-r+1)^+}(z) &=&  C_2(G) {1 \over z(1-z)^{2s-2r+1}},~~~~~
P^{(s-r+1)^+}_{r^+ s^+}(z) = C_2(G) {(1-z)^{2s-2r+3} \over z}, \nn\\
P^{r^-}_{s^+~ (s-r+1)^+}(z) &=&  C_2(G) {1  \over (1-z) z^{2s-2r+1}},~~~~~
P^{r^+}_{(s-r+1)^+~ s^+}(z) =  C_2(G) { z^{2s-2r+3}  \over (1-z)},
\eeqa
where $s\geq 2r -1$ and finally for
\beqa\label{lastindex}
A=r,~~~B=s-r-1,~~~~C=s~~~~~ \Rightarrow~~~~ n= (s-r-1),~~~m =- (s-r-1)
\eeqa
we get
\beqa\label{setofgluon2polarizeds-r-1}
P^{s^+}_{(s-r-1)^- ~r^+}(z) &=&  C_2(G) {z^{2s-2r-1} \over (1-z)^{2s-2r-3}},~~~~~
P^{(s-r-1)^-}_{s^+ r^+}(z) = C_2(G) {(1-z)^{2s-2r-1} \over z^{2s-2r-3}},\nn\\
P^{s^+}_{ r^- ~(s-r-1)^+}(z) &=&  C_2(G) {1 \over z(1-z)^{2s-2r-3}},~~~~~
P^{(s-r-1)^+}_{r^+ s^+}(z) = C_2(G) {(1-z)^{2s-2r-1} \over z}, \nn\\
P^{r^-}_{s^+~ (s-r-1)^+}(z) &=&  C_2(G) {1  \over (1-z) z^{2s-2r-3}},~~~~~
P^{r^+}_{(s-r-1)^+~ s^+}(z) =  C_2(G) { z^{2s-2r-1}  \over (1-z)},
\eeqa
where $s \geq 2r+1$. These general expressions
(\ref{setofgluon2polarizeds-r+1}), (\ref{setofgluon2polarizeds-r-1})
describe all possible splitting probabilities
corresponding to the interaction vertices with one space-time derivative and
therefore to the  dimensionless coupling constant, as it is in the generalized YM theory
\cite{Savvidy:2014hha,Savvidy:2005fi}.

\section{\it Generalization of DGLAP Equation}

Having in hand the new set of splitting probabilities
for tensor-gluons (\ref{setofgluon2polarizeds-r+1}), (\ref{setofgluon2polarizeds-r-1}),
one can calculate a possible emission of tensor-gluons which appears
in addition to the quark-anti-quark and gluon "clouds' in a proton.
Our goal is to derive
DGLAP equations \cite{Altarelli:1977zs,Gribov:1972ri,Gribov:1972rt,Lipatov:1974qm,Fadin:1975cb,Kuraev:1977fs,
Balitsky:1978ic,Dokshitzer:1977sg} which will take into account these new emission processes.

In accordance with our hypothesis there
is  an additional emission of tensor-gluons in a proton, therefore one should introduce
the corresponding densities $T_s(x, t)$ of tensor-gluons (summed over colors)
inside a proton in the $P_{\infty}$ frame \cite{Savvidy:2014hha}. We can  derive therefore
the integro-differential equations that describe the $Q^2$ dependence
of parton densities in this general case. They are:
\beqa\label{evolutionequation}
{d q^i(x,t)\over dt} &=& {\alpha(t) \over 2 \pi} \int^{1}_{x} {dy \over y}[\sum^{2 n_f}_{j=1} q^j(y,t)~
P_{q^i q^j}({x \over y})+ G(y,t)~ P_{q^i G}({x \over y})] ,\\
{d G(x,t)\over dt} &=& {\alpha(t) \over 2 \pi} \int^{1}_{x}
{dy \over y}[\sum^{2 n_f}_{j=1} q^j(y,t)~
P_{G q^j}({x \over y})+ G(y,t) ~P_{G G}({x \over y})+ \sum_{s} T_s(y,t) ~P_{G T_s}({x \over y}) ],\nn\\
{d T_r(x,t)\over dt} &=& {\alpha(t) \over 2 \pi} \int^{1}_{x} {dy \over y}[
G(y,t)~ P_{T_r G}({x \over y}) +  \sum_{s} T_{s}(y,t)~ P_{T_{r} T_{s} }({x \over y})].\nn
\eeqa
The $\alpha(t)$ is the running coupling constant ($\alpha = g^2/4\pi$).
In the leading logarithmic approximation $\alpha(t)$ is of the form
\be\label{strongcouplingcons}
{\alpha \over \alpha(t)} = 1 +b ~\alpha ~t~~,
\ee
where $\alpha = \alpha(0)$ and $b$ is the one-loop Callan-Simanzik coefficient,
which  receives an additional contribution from the tensor-gluons running in the loop.
Here the indices i and j run over quarks and antiquarks of all flavors and s and r run over
all spins of the tensor-gluons. The number
of quarks of a given fraction of momentum changes when a quark looses momentum by
radiating a  gluon
or a  gluon inside the proton produces a quark-antiquark pair \cite{Altarelli:1977zs}.
Similarly the number of  gluons changes
because a quark may radiate a gluon or because a gluon may split into a quark-antiquark
pair or into two gluons or {\it into two tensor-gluons}. The density of
tensor-gluons also changes because there are transitions between them through the
splittings which are described by the probabilities (\ref{setofgluon2polarizeds-r+1})
and (\ref{setofgluon2polarizeds-r-1}).

In order to guarantee that the total momentum of the proton, that is, of
all partons is unchanged, one should impose the following constraint:
\be\label{conservation}
{d\over dt}\int_{0}^{1} dz z [\sum^{2n_f}_{i=1}q^{i}(z,t)+G(z,t)+\sum_s T_s(z,t)]=0.
\ee
Using the evolution equations (\ref{evolutionequation}) one can express the derivatives
of the densities in (\ref{conservation}) in terms of splitting probabilities $P^C_{BA}$
(kernels of the evolution equations) and derive that the following
momentum sum rules should be fulfilled:
\beqa\label{momentumsum}
&&\int_{0}^{1} dz z [P_{qq}(z)+P_{Gq}(z) ]=0,\nn\\
&&\int_{0}^{1} dz z [2 n_f P_{qG}(z)+P_{GG}(z)+ \sum_s P_{T_s G}(z)]=0,\nn\\
&&\int_{0}^{1} dz z [ P_{GT_r}(z)+ \sum_s P_{T_sT_r}(z)]=0.
\eeqa
Before analyzing these momentum sum rules let us first
inspect the behavior of the tensor-gluon kernels (\ref{setoftensorgluonpolarized}),
(\ref{setofgluon2polarizeds-r+1}), (\ref{setofgluon2polarizeds-r-1})
at the end points $z=0,1$. As one can see,
they are singular at the boundary values similarly to the
case of the standard kernels (\ref{setofquarkgluon}).
Though there is a difference here: the singularities are of higher order compared to the standard case
\cite{Altarelli:1977zs}.
Therefore one should define the regularization procedure for the singular factors
$(1 - z)^{-2s+1}$ and $ z^{-2s+1}$  reinterpreting them as the  distributions $(1 - z)^{-2s+1}_{+}$ and
$z^{-2s+1}_{+}$ \cite{Savvidy:2014hha}, similarly to the Altarelli-Parisi regularization. The details
are given in the Appendix.

 The first equation
in the momentum sum rule (\ref{momentumsum})
remains unchanged because there is no tensor-gluon contribution into the quark evolution. The second
equation in the momentum sum rule (\ref{momentumsum}) will take the following form
(see \cite{Savvidy:2014hha} and Appendix for details):
\beqa\label{betacoefficient}
&\int_{0}^{1} dz z [2 n_f P_{qG}(z)+P_{GG}(z)+ \sum_s P_{T_s G}(z) + b_G \delta (z-1)]=\nn\\
&=\int_{0}^{1} dz z  [2 n_f T(R)[z^2 +(1-z)^2]+C_2(G) \left[{1 \over z(1-z)}
+ {z^4 \over z(1-z)}+{(1-z)^4 \over z(1-z)}\right]+\nn\\
&+C_2(G) \sum_s  \left[ {z^{2s+1}  \over (1-z)^{2s-1} }
+{(1-z)^{2s+1} \over z^{2s-1} }  \right]  ] +b_{11}=\nn\\
&={2\over 3} n_f T(R) - {11 \over 6}C_2(G)- \sum_s {12 s^2 -1\over 6} C_2(G) + b_{11} =0.
\eeqa
We can extract now the additional contribution
to the one-loop Callan-Symanzik beta function
for gluons $b_{11} $ arising from the tensor-gluon loop of spin s \cite{Savvidy:2014hha}:
\be\label{spinscontr}
b_{11} =  \sum_s {12 s^2 -1\over 6} C_2(G), ~~~s=1,2,3,4,....
\ee
At s=1 we are rediscovering the
asymptotic freedom result \cite{Gross:1973ju,Gross:1974cs,Politzer:1973fx}.
For larger spins the  tensor-gluon contribution into the Callan-Simanzik beta function
has the same signature as the standard gluons, which means that tensor-gluons
"accelerate" the asymptotic freedom (\ref{strongcouplingcons}) of the strong
interaction coupling constant $\alpha(t)$.
At large transfer momentum
the strong coupling constant tends to zero faster compared to the standard case
\cite{Savvidy:2014hha}\footnote{In (\ref{spinscontr}) one should take into account the
multiplicity of higher spin states in the theory under consideration.}.

One can confirm the above result by using effective action approach developed in the Yang-Mills
theory \cite{Savvidy:1977as,Matinyan:1976mp,Batalin:1976uv,Kay:1983mh} extended
to the higher spin gauge bosons. With the spectrum of the tensor-gluons in
the external chromomagnetic field  $k^2_0 = (2n+1 + 2s)gH +k^2_{\perp}$
one can perform a summation of the modes and get an exact result for the one-loop effective
action similar to \cite{Savvidy:1977as,Kay:1983mh}:
\be
\epsilon= {H^2 \over 2} +{(gH)^2 \over 4\pi} ~b_{11} ~[\ln{gH \over \mu^2}-{1\over 2}],
\ee
where
\beqa\label{savv}
b_{11} =  -{2 C_2(G)\over \pi} ~ \zeta(-1, {2s+1\over 2})=
{12s^2-1\over 12 \pi}C_2(G),
\eeqa
and $\zeta(-1, q)=-{1\over 2}(q^2 -q +{1\over 6})$ is the generalized zeta
function\footnote{The generalized zeta function is defined as $\zeta(p, q)=\sum^{\infty}_{k=0}{1\over (k+q)^p}
={1\over \Gamma(p)} \int^{\infty}_{0} dt t^{-1+p} { e^{-qt} \over 1-e^{-t}} $ .}.
Because the coefficient in front of the logarithm defines  the beta function
\cite{Savvidy:1977as,Matinyan:1976mp}, one can see that (\ref{savv}) is
in agreement with the result (\ref{spinscontr}).

The third equation in the momentum sum rule (\ref{momentumsum}) will take the following form:
\beqa\label{tensorbeta}
&\int_{0}^{1} dz z [P_{GT_r}(z) + \sum_s P_{T_s T_r}(z) +   b_{rr} \delta (z-1)]=0,
\eeqa
and we shall get
\beqa\label{tensorbetar}
& C_2(G)\int_{0}^{1} dz z
 \sum_{s} [{z^{2s-2r+3} \over  (1- z)^{2s-2r+1}} +{(1-z)^{2s -2r+3} \over z^{2s-2r+1}}]
+ b_{rr}  = \nn\\
&= -\sum_{s} {12 (s-r+1)^2 -1\over 6} C_2(G)  + b_{rr} =0.
\eeqa
We can extract  the one-loop coefficient of the Callan-Symanzik beta function now for
tensor-gluon of spin r, which has the form
\be\label{callansimanzicgeneral}
b_{rr}  = \sum_s {12 (s-r+1)^2 -1\over 6} C_2(G), ~~~r=1,2,3,4,....,
\ee
and it has the identical, {\it quadratic}, dependence on the particle spins
running in the loop, as it is for gluons (\ref{spinscontr}).
The consistency relation now will take the following form:
\be\label{consistancyequation}
b_{11} =  \sum_s {12 s^2 -1\over 6} C_2(G) = \sum_{s^{'}} {12 (s^{'}-r+1)^2 -1\over 6} C_2(G)=b_{rr}.
\ee
In order to perform the summation over spins in the above equation  one should take into account
the multiplicity of higher spin states in the theory under consideration. In the generalized YM theory
the values of the tensor-gluon spins run to infinity and the multiplicity is given in
\cite{Savvidy:2010vb}:
\beqa\label{spectrum}
~~~~~~~~~~\pm(s-1),~~~~~\pm(s-3),...\nn\\
\pm (s+1),~~~~~~~~~~~~~~~~~~~~~~~~~~~~~~~~~~~~~~~~\\
~~~~~~~~~~\pm (s-1),~~~~~\pm(s-3),...\nn
\eeqa
Therefore the sums on the both sides of the equation (\ref{consistancyequation})
 are diverging. One can suggest two scenarios \cite{Savvidy:2014hha}.
In the first one the high spin gluons, let us say of $s  \geq 3$, will get large mass
and therefore can be ignored at a given energy scale. In the second case,
when all of them are massless or what is the same the energy is much larges than
all the masses of the tensor-gluons, then one can
suggest the Reimann zeta function regularization, similar to the Brink-Nielsen regularization
\cite{Brink:1973kn,Bakas:2004jq}. The spectrum (\ref{spectrum}) is represented in the following way:
\beqa
&&\pm 1\nn\\
&&\pm 2, ~~~~~0\nn\\
&&\pm 3,~~\pm 1,~~\pm 1\nn\\
&&\pm 4,~~\pm 2,~~\pm 2,~~~~0\nn\\
&&\pm 5,~~\pm 3,~~\pm 3,~~\pm 1,~~\pm 1  \nn\\
&&\pm 6,~~\pm 4,~~\pm 4,~~\pm 2,~~\pm 2,~~~~0\nn\\
&&....................................................,
\eeqa
and the summation over columns gives:
\beqa
b_{11} &=& C_2(G) [ \sum^{\infty}_{s=1} {( 12s^2 -1) \over 12 \pi}+
\sum^{\infty}_{s=0} {( 12s^2 -1) \over 12 \pi} + \sum^{\infty}_{s=1} {( 12s^2 -1) \over 12 \pi}+
\sum^{\infty}_{s=0} {( 12s^2 -1) \over 12 \pi} +.....] =\\
&=&C_2(G)[{1\over \pi}  \zeta(-2)- {1\over 12\pi} \zeta(0)-{1\over 12\pi}
+{1\over \pi}  \zeta(-2)- {1\over 12\pi} \zeta(0) +...] =C_2(G)[{1\over 24\pi} - {1\over 12\pi} +
{1\over 24\pi}+...]=0,\nn
\eeqa
where $\zeta(-2)=0,~ \zeta(0)=-1/2$, leading to the theory which is  {\it conformally invariant} at very high energies. The above
summation requires explicit regularisation and further justification.

\section{\it Conclusion}

In this article we continue to analyze  a possibility that inside a proton and, more generally,
inside hadrons there are additional partons - tensor-gluons, which can carry a part of the proton
momentum \cite{Savvidy:2014hha}.   A nonzero density of the tensor-gluons can be generated
by the emission of tensor-gluons by gluons.  The tensor-gluons themselves further split into
the pairs of tensor-gluons through different channels. Therefore the density of neutral partons in a proton is
given by the sum:
$$
G(x,t)+\sum_s T_s(x,t),
$$
where $T_s(x,t)$ is the density of the
tensor-gluons of spin s. To describe these processes between gluons and tensor-gluons one
should know the general splitting probabilities for tensor-gluons $P^C_{BA}$. These probabilities
fulfill very general symmetry relations, which we were able to
resolve  by introducing a splitting index (\ref{index}).  This approach allows to find out
the general form of the splitting  functions (\ref{setofgluon2polarizeds-r+1}),
(\ref{setofgluon2polarizeds-r-1}), to derive corresponding DGLAP evolution
equations (\ref{evolutionequation}) and to calculate the one-loop Callan-Simanzik beta function
for tensor-gluons of different spins (\ref{callansimanzicgeneral}).  Our results provide
a nontrivial consistency check of the theory and of the  Callan-Simanzik beta function calculations,
because the theory has a unique coupling constant and its high energy behavior should be universal
for all particles of the spectrum. We argue that the contribution of all spins into the beta function  vanishes leading to a conformal invariant theory at very high energies.

This work was supported in part by the
General Secretariat for Research and Technology of Greece and
from the European Regional Development Fund MIS-448332-ORASY (NSRF 2007-13 ACTION, KRIPIS).

\section{\it Note Added}
One can suggest an  alternative approach to calculating the splitting functions which 
is based on deformation of the momenta of the
particles in the triple vertices   (\ref{rssvertex}) by a complex parameter w  without breaking 
the mass shell conditions \cite{Antoniadis:2011rr} 
\be\label{momenta}
p_1=(\omega z, w, iw, k z),~p_2=(\omega (1-z), -w, -iw, k (1-z)),~(\omega , 0, 0, k ).
\ee
The corresponding polarization vectors are to be taken in the form:
\be
e^+_1 = {1\over \sqrt{2} } ( {z \over \omega}, 1, -i, - {z \over k}),~
e^+_2 = {1\over \sqrt{2} } (- {z \over \omega}, 1, -i, - {z \over k}),~
e^-_3 = {1\over \sqrt{2} } ( 0, 1, -i, 0),
\ee
so that to fulfill the following relations 
\be
p^2_1=p^2_2=p^2_3=p_1 p_2=p_2 p_3=p_3 p_1=p_1 e^+_1=p_2 e^+_2=p_3 e^-_3=0.
\ee
The spinor representation of the momenta (\ref{momenta}) will take the form
\beqa
&\lambda_1=(\sqrt{\omega+k)z},0),~~~&\tilde{\lambda}_{\dot{1}}=(\sqrt{\omega+k)z},{2w\over \sqrt{\omega+k)z}}),\nn\\
&\lambda_2=(\sqrt{\omega+k)(1-z)},0),~~~&\tilde{\lambda}_{\dot{2}}=(\sqrt{\omega+k)(1-z)},-{2w\over \sqrt{\omega+k)(1-z)}}),\nn\\
&\lambda_3=(\sqrt{\omega+k)},0),~~~&\tilde{\lambda}_{\dot{3}}=(\sqrt{\omega+k)},0).
\eeqa
One can see that the invariant products
$
<1,2>=<2,3>=<3,1>=0
$
vanish and that 
\be\label{products}
[1,2]=-2w {1\over \sqrt{z(1-z)}},~~~[2,3]=2w {1\over \sqrt{1-z}},~~~[3,1]=2w {1\over \sqrt{z}}.
\ee
Let us consider the interaction vertices (\ref{rssvertex}) which include the tensor-gluons
of spin $A=r$ (helicities $h_2=\pm r$), of spin $C=s$ (helicities $h_1 = \pm s$) and 
the third spin of the type:
$B =  s-r+1 $, $s \geq 2r-1$,  where $r=1,2,3...$. A straightforward  calculation with the use of 
scalar products (\ref{products}) gives:
\beqa\label{rssvertex11}
M_3(1^{-s} ,2^{+r},3^{s-r+1} )&\propto&  { [2,3]^{2s+1} \over [1,2]^{2s-2r+1}  [3,1]^{2r-1}}=
-2 w {z^s \over (1-z)^r}\nn\\
M_3(1^{-s} ,2^{s-r+1},3^{+r} )&\propto&  {[2,3]^{2s+1} \over [1,2]^{2r-1}  [3,1]^{2s-2r+1}}=
-2 w {z^s \over (1-z)^{s -r+1} }\nn\\
M_3(1^{+r} ,2^{s-r+1},3^{-s} )&\propto&  {[1,2]^{2s+1} \over [2,3]^{2r-1}  [3,1]^{2s-2r+1}}=
-2 w {1 \over z^r (1-z)^{s -r+1} }\nn\\
M_3(1^{+r} ,2^{-s},3^{s-r+1} )&\propto&  {[3,1]^{2s+1} \over [2,3]^{2r-1}  [1,2]^{2s-2r+1}}=
-2 w {(1-z)^{s} \over z^r  }\nn\\
M_3(1^{s-r+1} ,2^{+r},3^{-s} )&\propto&  {[1,2]^{2s+1} \over [3,1]^{2r-1}  [2,3]^{2s-2r+1}}=
-2 w {1 \over z^{s-r+1} (1-z)^{r} }\nn\\
M_3(1^{s-r+1} ,2^{-s},3^{+r} )&\propto&  {[3,1]^{2s+1} \over [1,2]^{2r-1}  [2,3]^{2s-2r+1}}=
-2 w {(1-z)^{s} \over z^{s-r+1}  }.
\eeqa
Considering the transversal momentum $p_{\perp}$  in (\ref{transverce})
to be proportional to the  deformation parameter $p_{\perp} \propto w$   
one can get the following expression  for splitting  probabilities
\beqa
P(z)= {1\over 2}z(1-z) \vert M_3 \vert^2 {1\over \vert w \vert^2} 
\eeqa
and then, using (\ref{rssvertex11}), the following set of splitting probabilities 
\beqa
P^{s}_{s-r+1,r} &=& C_2(G) {(1-z)^{2s+1} \over z^{2s-2r+1}  },~~~~
P^{s-r+1}_{s,r} = C_2(G)  {z^{2s+1} \over (1-z)^{2s -2r+1} }\nn\\
P^{s}_{r,s-r+1} &=& C_2(G) {(1-z)^{2s+1} \over z^{2r-1}  },~~~~
P^{s-r+1}_{r,s} =  C_2(G){1 \over z^{2r-1} (1-z)^{2s -2r+1} }\nn\\	
P^{r}_{s,s-r+1}&=& C_2(G)  { z^{2s+1} \over (1-z)^{2r-1}   },~~~~
P^{r}_{s-r+1,s} = C_2(G) {1 \over z^{2s-2r+1} (1-z)^{2r-1} },
\eeqa
where $s> 2r-1$.  With this splitting probabilities the indices (\ref{index}) are equal   to
\be
A=r,~~~B=s-r-1,~~~~C=s~~~~~ \Rightarrow~~~~~~~n=+s,~ m= -(s-r+1)
\ee
and differ from the {\it conjectured} value for the index $n=s-r+1$ in (\ref{lastindex}).   
The splitting probabilities of spin $A=r$, of spin $C=s$  and  $B =  s-r-1 $ are 
\beqa
P^{s}_{s-r-1,r} &=& C_2(G) {  z^{2s-2r-1}  \over  (1-z)^{2s-1}},~~~~
P^{s-r-1}_{s,r} = C_2(G)  {(1-z)^{2s -2r-1} \over  z^{2s-1} }\nn\\
P^{s}_{r,s-r-1} &=& C_2(G) {z^{2r+1}  \over (1-z)^{2s-1}  },~~~~
P^{s-r-1}_{r,s} =  C_2(G) z^{2r+1} (1-z)^{2s -2r-1} \nn\\	
P^{r}_{s,s-r-1}&=& C_2(G)  { (1-z)^{2r+1}   \over  z^{2s-1} },~~~~
P^{r}_{s-r-1,s} = C_2(G)   z^{2s-2r-1} (1-z)^{2r+1} ,
\eeqa
where $s \geq 2r+1$  $r=1,2,3...$ .

The momentum sum rule (\ref{tensorbeta}) will take the following form:
\beqa\label{tensorbetar1}
& C_2(G)\int_{0}^{1} dz z
 \sum_{s} [{z^{2s+1} \over  (1- z)^{2s-2r+1}} +{(1-z)^{2s +1} \over z^{2s-2r+1}}]
+ b_{rr}  = \nn\\
&= C_2(G)\sum^{2s+1}_{k=2s-2r+1} {{(2s+1)!} \over k! (2s+1-k)!}{1\over k-2s+2r}  + b_{rr} =0,
\eeqa
and one  can extract  the one-loop  Callan-Symanzik beta function  for
the tensor-gluon of spin r:
\be\label{callansimanzicgeneral1}
b_{rr}  = -C_2(G) \sum_s \sum^{2s+1}_{k=2s+1-2r} {{(2s+1)!} \over k! (2s+1-k)!}{1\over k-2s+2r}, ~~~r=1,2,3,4,....,
\ee
In particular, for spin $r=2$ one can get
\beqa
b_{22}  = -C_2(G) \sum_s \sum^{2s+1}_{k=2s-3} {{(2s+1)!} \over k! (2s+1-k)!}{1\over k-2s+4}
=C_2(G)\sum_{s=3}  {4 s^4 -8s^3 +3s^2 +s -3/10\over 6}, 
\eeqa 
which grows as $s^4$ compared to the  $s^2$ in (\ref{consistancyequation}).
In this respect a further study of the splitting probabilities is required. I would like 
to thank G.Georgiou and S.Konitopoulos for discussion which leades to this 
note.

\section{\it Appendix}

The regularisation is defined in the following
way \cite{Savvidy:2014hha}:
\beqa\label{definition}
\int_{0}^{1} dz {f(z)\over (1 - z)^{2s-1}_+}&=&
\int_{0}^{1} dz {f(z)- \sum^{2s-2}_{k=0} {(-1)^k \over k!} f^{(k)}(1) (1-z)^k \over (1 - z)^{2s-1}},\nn\\
\nn\\
\int_{0}^{1} dz {f(z)\over z ^{2s-1}_+}&=&
\int_{0}^{1} dz {f(z)- \sum^{2s-2}_{k=0} {1 \over k!} f^{(k)}(0) z^k \over z^{2s-1}},\\
\nn\\
\int_{0}^{1} dz {f(z)\over z_+ (1-z)_+}&=&
\int_{0}^{1} dz {f(z)-  (1-z)f(0) - z f(1) \over z  (1-z) },\nn
\eeqa
where $f(z)$ is any test function which is sufficiently regular at the end points
and, as one can see, the defined substraction guarantees the convergence of the integrals.
Using the same arguments as in the standard case \cite{Altarelli:1977zs} we should add the delta function
terms into the definition of the  kernels.

The momentum sum rule integrals (\ref{momentumsum}), (\ref{betacoefficient})
and (\ref{tensorbeta}) can be calculated in two different ways: with direct regularization
of the integrals  near $z=1$ and near $z=0$, as it is defined in (\ref{definition}),
or by using the substitution $w=1-z$ in the second term of the integrand and then
regularizing  the resulting integrand near $z=1$. As one can get convinced both methods
give identical results. Therefore  the integral over
$P_{TG}(z)$ in (\ref{betacoefficient}) gives:
\beqa
 & \int_{0}^{1} dz   \left[ {z^{2s+2} \over (1-z)^{2s-1}_+}
+{(1-z)^{2s+1} \over z^{2s-2}_+ }  \right]   =\nn\\
& = \int_{0}^{1} dz \left[
  \sum^{2s+2}_{k=2s-1}   {  (-1)^k (2s+2)!  \over k!(2s+2 -k)! }~
{(1-z)^k \over  (1-z)^{2s-1}}  +
 \sum^{2s+1}_{k=2s-2}   {  (-1)^k (2s+1)!  \over k!(2s+1 -k)! }~
{z^k \over  z^{2s-2}}  \right]  =\nn\\
&= - {12 s^2 -1\over 6}  .
\eeqa
The same integral  can be calculated by the substitution $w=1-z$ in the
second term of the integrand
\beqa
&  \int_{0}^{1} dz    {z^{2s+1} \over (1-z)^{2s-1}_+}=  \int_{0}^{1} dz
  \sum^{2s+1}_{k=2s-1}  \left[ {  (-1)^k (2s+1)!  \over k!(2s+1 -k)! }~
{(1-z)^k \over  (1-z)^{2s-1}}  \right]  = - {12 s^2 -1\over 6}  ,\nn
\eeqa
which is identical to the previous result.

\end{document}